\documentclass[12pt]{article}
\usepackage{amsmath,amssymb,amsfonts}
\usepackage[paper=letterpaper,margin=1.0in]{geometry}

\newcommand{\be}{\begin{equation}}
\newcommand{\ee}{\end{equation}}
\newcommand{\bea}{\begin{eqnarray}}
\newcommand{\eea}{\end{eqnarray}}
\newcommand{\ba}{\begin{array}}
\newcommand{\ea}{\end{array}}

\newcommand{\BC}{{\mathbb C}}

\newcommand{\CP}{{\mathbb C}{\mathbb P}}

\newcommand{\Diff}{{\rm Diff}}
\newcommand{\Gr}{{\rm Gr}}

\newcommand{\tr}[1]{{\rm tr}\, {1}}
\newcommand{\bra}[1]{\langle{1}|}
\newcommand{\ket}[1]{|{1}\rangle}
\newcommand{\ip}[2]{\langle{1}|{2}\rangle}
\newcommand{\vev}[1]{\langle{1}\rangle}

\newcommand{\todo}[1]{{\bf {1}}}
\newcommand{\comment}[1]{}

\def\dd{{\rm d}}

\newcommand{\ft}[2]{{\textstyle\frac{1}{2}}}
\def\fract12{{\textstyle{1\over2}}}
\def\ffract12{\raise .3 em\hbox{$\scriptstyle1$}\kern-.25em/
             \kern-.2em\lower .2 em \hbox{$\scriptstyle2$}}
\def\fractje12{{\scriptstyle{1\over2}}}

\def\part12{{\partial1\over\partial2}}
\def\ex1{e^{\textstyle1}}

\begin{document}

{\footnotesize
\rightline{IH\'ES P/09/23}
\rightline{VPI-IPNAS-09-04}
}

\renewcommand{\thefootnote}{\fnsymbol{footnote}}
\vskip1.5cm
\centerline{\Large \bf The Big Bang as the Ultimate Traffic Jam}
\vskip1.5cm

\centerline{{\bf
Vishnu Jejjala${}^{1}$\footnote{\tt vishnu@ihes.fr},
Michael Kavic${}^{2}$\footnote{\tt kavic@vt.edu},
Djordje Minic${}^{2}$\footnote{\tt dminic@vt.edu},
Chia-Hsiung Tze${}^{2}$\footnote{\tt kahong@vt.edu}}}
\vskip0.5cm
\centerline{${}^1$\it Institut des Hautes \'Etudes Scientifiques}
\centerline{\it 35, Route de Chartres}
\centerline{\it 91440 Bures-sur-Yvette, France}
\vskip0.5cm
\centerline{${}^2$\it Institute for Particle, Nuclear, and Astronomical Sciences}
\centerline{\it Department of Physics, Virginia Tech}
\centerline{\it Blacksburg, VA 24061, U.S.A.}
\vskip0.5cm

\begin{abstract}
We present a novel solution to the nature and formation of the initial state of the Universe.
It derives from the physics of a generally covariant extension of Matrix theory.
We focus on the dynamical state space of this background independent quantum theory of gravity and matter, an infinite dimensional, complex non-linear Grassmannian.
When this space is endowed with a Fubini--Study-like metric, the associated geodesic distance between any two of its points is zero.
This striking mathematical result translates into a physical description of a hot, zero entropy Big Bang.
The latter is then seen as a far from equilibrium, large fluctuation driven, metastable ordered transition, a ``freezing by heating'' jamming transition.
Moreover, the subsequent unjamming transition could provide a mechanism for inflation while rejamming may model a Big Crunch, the final state of gravitational collapse.\footnote{
This essay received an honorable mention in the Gravity Research Foundation Essay Competition, 2009. 

}
\end{abstract}

\setcounter{footnote}{0}
\renewcommand{\thefootnote}{\arabic{footnote}}

\newpage

\section{Where to begin?}
Explanations for how the Universe began are as ancient as civilization, but it is only in the last century that we have been able to commence a rigorous and scientific examination of the question in light of observational data.
The observed expansion of the Universe together with measurements of the cosmic microwave background radiation (CMBR) vindicate the paradigm of a hot Big Bang~\cite{wmap}.
Standard cosmological models propose an initial spacelike singularity.
Such a state signals the breakdown of spacetime and geometry as effective descriptions of Nature.
Understanding the physics of the singularity and the dynamical evolution of the Universe at the earliest times remains one of the long standing and unrealized ambitions of any putative quantum theory of gravity.

Running time's arrow in reverse, the second law of thermodynamics tells us that the initial condition for the Universe must have had a very low entropy~\cite{penrose}.
Indeed, from the viewpoint of the {\em unique} Hartle--Hawking wave function, this entropy ought to be {\em zero}.
Yet accounting for a zero entropy initial condition remains a fundamental puzzle for cosmology.

Another mystery follows.
While inflationary models successfully resolve the horizon, flatness, and magnetic monopole problems and are in accord with precision measurements of the CMBR, their underlying mechanism is at best unclear.
As the dynamical process closest in time to the Big Bang event, could inflation be a natural or inevitable consequence of how the Universe began?
We shall argue in this essay that the resolution of the cosmological singularity could in fact facilitate the inflationary phase of expansion in the early Universe.


We address the origin of the Universe, specifically the very nature of its initial state, in the context of a previously proposed generalization of quantum theory, a background independent formulation of Matrix theory~\cite{mt, dj, review}.
A key feature of this extension is a new state space given by an infinite dimensional, complex non-linear Grassmannian.
It is non-linear because it is the coset space of diffeomorphism groups.
In fact, it is the natural diffeomorphism invariant generalization of the linear Grassmannian $\CP^n$ of ordinary quantum mechanics.
For background purposes, we provide only a conceptual summary of how this situation comes about;
details may be found in the review~\cite{review}.

Although the geometric formulation may be unfamiliar, standard quantum mechanics can be cast as Hamiltonian dynamics over a specific phase space, $\CP^n$, the complex projective Hilbert space of pure states~\cite{review, geoqm}.
$\CP^n$ is a K\"ahler--Einstein manifold with constant, holomorphic sectional curvature $\frac2\hbar$.
Being K\"ahler, it possesses notably a triad of compatible structures, any two of which determine the third.
These are a symplectic two-form $\omega$, an unique Fubini--Study (FS) metric $g$, and a complex structure $j$.
All the key features of quantum mechanics are encoded in the geometric structure of $\CP^n$.
In particular, the Riemannian metric determines the distance between states on the phase space.
The geodesic distance is a measure of change in the system, for example through Hamiltonian time evolution.
By way of the FS metric and the energy dispersion $\Delta E$, the infinitesimal distance in phase space is $\dd s = \frac{2}{\hbar}\ \Delta E\ \dd t$.

Through this relation, time reveals its statistical, quantum nature.
It also suggests that dynamics in time relate to the behavior of the metric on the configuration space.
The Schr\"odinger equation is simply a geodesic-like equation for a particle moving on $\CP^n = U(n+1)/(U(n)\times U(1))$ in the presence of an effective external $U(n)\times U(1)$ gauge field whose source is the Hamiltonian of a given physical system.
This geometric non-linearity of $\CP^n$ and of the geodesic-like Schr\"odinger equation become linear since these structures can be lifted into the linear Hilbert space as is familiar in the usual formulation of quantum theory where only $\hbar$ and the dimension $n$ (generally infinite) of the space remain.
That classical physical space may be {\em emergent} is hinted at by a simple observation.
When the configuration space of the system is the {\em physical} space, the unique FS metric reduces precisely to the {\em spatial} metric.
If we consider extending the geometry of quantum mechanics to induce emergent, arbitrarily curved spacetimes, as would be expected in a theory of quantum gravity, we should generalize the space of states and also its metric structure.

Matrix theory, we recall, is a manifestly second quantized, non-perturbative formulation of M-theory on a fixed spacetime background~\cite{bfss}.
To construct a background independent formulation of Matrix theory, we relax, albeit slightly, the rigidity of the underlying quantum theory.
Our minimal extension is done by way of a new symmetry, a quantum diffeomorphism invariance.
We require the following features~\cite{mt, review, jm}:
At the basic level, there are only dynamical correlations between quantum events.
The phase space must have a symplectic structure, namely a symplectic two-form;
it must be diffeomorphism invariant;
and it must be the base space of a $U(1)$ bundle --- {\em i.e.}, there is a Berry phase.
We demand a three-way interlocking of the Riemannian, the symplectic, and the non-integrable almost complex structures.

In so departing from the integrable complex structure of $\CP^n$, the quantum mechanical phase space is given by the non-linear Grassmannian,
$\Gr(\BC^{n+1}) = \Diff(\BC^{n+1})/\Diff(\BC^{n+1}, \BC^n\times \{0\})$,
with $n\to\infty$, a complex projective, strictly almost K\"ahler manifold.
This last, strictly almost complex property has the physical interpretation of promoting the global time of quantum mechanics to a more provincial local time, a clearly desirable feature for classical and quantum gravity.
Note that we can now generalize the line element so that the energy uncertainty is measured in terms of a fundamental energy scale, the Planck energy, $E_{P}$, so that $\dd s = \frac{2}{\hbar}\, E_P\, \dd t$.
When applied to a cosmological setting, this relation defines cosmological time;
it is interesting to note that setting $\dd t=0$ implies $\dd s = 0$.

Most importantly, the property of diffeomorphism invariance implies that not just the metric but also the almost complex structure and hence the symplectic structure must be fully dynamical.
Consequently, with the coadjoint orbit nature of $\Gr(\BC^{n+1})$, the equations of motion of this general theory are Einstein-like equations with the energy-momentum tensor being determined by a holonomic Yang--Mills field strength, the Hamiltonian (``charge''), and a ``cosmological constant'' term.
We shall refer to these as the Einstein--Yang--Mills equations over state space.
Just as the geodesic equation for a non-Abelian charged particle is contained in the classical Einstein--Yang--Mills equations, so is the corresponding geometric, covariant Schr\"odinger equation.

\comment{
\be
\label{BIQM1}
{\cal{R}}_{ab} - \frac{1}{2} {\cal{G}}_{ab} {\cal{R}}  - \lambda {\cal{G}}_{ab}= {\cal{T}}_{ab} (H, F_{ab}) ~,
\ee
with ${\cal{T}}_{ab}$ as determined by ${\cal{F}}_{ab}$, the holonomic Yang--Mills field strength, the Hamiltonian (``charge'') $H$, and a ``cosmological'' term $\lambda$.
Furthermore,
\be
\label{BIQM2}
\nabla_a {\cal{F}}^{ab} = \frac{1}{2M_P} H u^b ~,
\ee
where $u^b$ are the velocities, $M_P$ is the Planck energy, and $H$ the Matrix theory Hamiltonian~\cite{bfss}.
These coupled equations imply via the Bianchi identity a conserved energy-momentum tensor: $\nabla_a {\cal{T}}^{ab} =0$.
}

Thus $\Gr(\BC^{n+1})$ has the necessary properties to be the phase space of quantum gravity, which itself underlies all of spacetime.
If this is correct, then it is reasonable to expect that this space should provide a description of manifestly quantum gravitational systems, including the initial cosmological singularity.
It is in fact the case that recent mathematical findings concerning this space indicate that it possesses precisely the correct properties to provide a description of the beginning of our Universe.
Moreover, the features of the non-linear Grassmannian suggest that the early Universe should be viewed as a far from equilibrium system that is jam packed like a freeway at rush hour.

\section{The Universe vanished in the most unlikely way}
In geometric quantum mechanics, every statement about the state space $\CP^n$ translates into a physical statement about quantum mechanics.
Similarly, we thereby expect a correspondence between the non-linear Grassmannian $\Gr(\BC^{n+1})$ and physics in our extended quantum theory.
Thus we will take seriously any mathematical property of the space and interpret it in a physical language.
One such attribute arises due to a remarkable theorem of Michor and Mumford~\cite{mm}.
It states that, as measured by the exact analogue of the FS metric on $\CP^n$, the geodesic distance between any two points on the non-linear Grassmannian vanishes.
Since $\Gr(\BC^{n+1})$ is the space of states out of which spacetime emerges, we see that the vanishing theorem naturally describes an initial state in which the Universe exists at a single point, the cosmological singularity.


Making use of the Michor--Mumford vanishing theorem, the low entropy problem tied to the initial conditions of the Universe is naturally resolved.
In the language of statistical geometry and quantum distinguishability, the generalized FS metric having vanishing geodesic distance means that none of the states of our non-linear Grassmannian phase space can be differentiated from each other.
Due to the large fluctuations in curvatures everywhere, the entire phase space is composed of a single, {\em unique} microstate.

A point in the phase space of the theory corresponds to a state of the quantum system under consideration, which in this case is the entire Universe.
This fact should be interpreted in a non-equilibrium setting.
Even though equilibrium thermodynamics does not hold, entropy continues to be a measure of the volume of the configuration space of the system.
Because the FS metric implies that points on $\Gr(\BC^{n+1})$ cannot be distinguished, we may infer via Boltzmann's formula~\cite{gs} that the entropy of the Universe is identically zero.
This is precisely the type of configuration that describes the initial state of the Universe.

\section{Everywhere was going nowhere fast}
What could be the physics of the low (zero) entropy, but high temperature state associated to the Big Bang?
Namely, how do we obtain an ordered state by heating?
We suggest that the paradoxical zero geodesic distance, everywhere high curvature property of $\Gr(\BC^{n+1})$ with the FS metric finds an equally paradoxical physical realization in the context of the model.
The physical interpretation is to be found in a class of far from equilibrium collective phase transitions, the so called ``freezing by heating'' jamming transitions.
A non-equilibrium state is required since in equilibrium, by heating, one necessarily increases the disorder of a system.
From many investigations~\cite{pettini}, it has been established that high curvatures in the phase or configuration manifold of a physical system precisely reflect large fluctuations of the relevant physical observables at a phase transition point.
This correspondence means equating the high curvatures of the FS metric on $\Gr(\BC^{n+1})$ with large (gravitational) fluctuations in our system at the phase transition.
In fact, the vanishing geodesic distance can be taken as the signature, or order parameter, of a strong fluctuation (or ``heat'') induced zero entropy and hence highly ordered state.

While from an equilibrium physics perspective such a state seems nonsensical, it does, however, occur in certain far from equilibrium environments.
Specifically, we point to a representative continuum model~\cite{helbing, zia, edwards} where such an unexpected state was first discovered.
Here, one has a system of particles interacting, not only through frictional and short range repulsive forces, but also and most importantly via strong driving fluctuations ({\em e.g.}, noise, heat, etc.).
As the amplitude of the fluctuations ({\em e.g.}, temperature) goes from weak to strong to extremely strong and as its total energy increases, such a system exhibits a thermodynamically counterintuitive evolution:
it goes from a fluid to a solid --- hence, a highly ordered, low entropy, crystalline metastable state --- and at last to a gas.

While the non-linear dynamics on $\Gr(\BC^{n+1})$ are far more intricate than in the above examples, which are more typically used to model traffic jams, they nevertheless have the requisite combination of the proper kind of forces to achieve these ``freezing by heating'' transitions.
Specifically, the system being considered here is far from equilibrium with low entropy, high temperature, and negative specific heat.
Its non-linear dynamics involve attractive and repulsive Yang--Mills forces, short range repulsive forces of D$0$-branes in the Matrix theory, repulsive forces from a positive ``cosmological'' term, and most importantly large gravitational fluctuations reflected in the large curvatures of the state space.

\enlargethispage\baselineskip
Now, it is known that geometric quantum mechanics can be seen as a classical completely integrable Hamiltonian system~\cite{block}, one with a K\"ahler phase space, a property tied in a one-to-one manner to the Hermiticity of all observables in their operatorial representations.
Similarly, our extended quantum theory is viewed as classical non-linear field and particle dynamics over a strictly almost complex phase space.
This last property implies that the corresponding operators would be non-Hermitian, and thereby our system is dissipative~\cite{rajeev}, as required to generate ``freezing by heating'' transitions.
Being a ``classical'' Einstein--Yang--Mills system, it is also non-integrable and chaotic~\cite{barrow}.

From the previous considerations, the following qualitative picture emerges.
From the relation between geodesic distance and time, we have the emergence of a cosmological arrow of time.
While the system has entropy $S=0$, the very high curvatures in $\Gr(\BC^{n+1})$ signal a non-equilibrium condition of dynamical instability.
Because of its non-linear dissipative and chaotic dynamics, our system will flow toward differentiation, which thereby yields, through entropy production, distinguishable states in the configuration space.
The dynamical evolution according to the second law is toward some higher entropy but stable state, the low-temperature Universe that we observe and inhabit today.

This scenario can occur as $\Gr(\BC^{n+1})$ has, in principle, an infinite number of metrics, a subset of which will solve the dynamical equations on the state space.
In fact, there is an infinite one parameter family of non-zero geodesic distance metrics, of which the FS metric is a special case~\cite{mm}.
During the dynamical evolution to a higher entropy state, spacetime and canonical quantum mechanics 
should emerge.

While the unjamming process is still not well understood in the non-equilibrium literature, we can, in a more speculative vein, envision its mapping within our context into an emergent spacetime setting.
Then the evolution away from a metastable jammed state to a stable equilibrium state, with its known universal scaling laws and its constituents' velocities potentially growing exponentially in time, would translate onto a cosmological evolution with its requisite period of inflationary exponential expansion followed by a graceful exit.
Specifically, consider the time derivative of the size of the Universe $a(t)$ in ${\rm d}s^2 = -{\rm d}t^2 + a^2(t)\, {\rm d}\Omega^2$, as being analogous to the velocity.
Long time tails in the velocity distribution could correspond to a Friedmann--Robertson--Walker (FRW) phase, in which there is power law behavior for $a(t)$.
Finally, jammed states, like many non-equilibrium phenomena, tend to self-tune their parameters, which might explain the slow-roll inflationary parameters in a natural, dynamical fashion.
Such an attractive scenario is currently under investigation.

\section{The end of the road}
To the above, we wish to add a related path to be explored.
The singularity at the heart of a gravitational collapse should encode an enormous number of degrees of freedom, of order $10^{{10^{80}}}$.
From our geometric perspective, this is related to the evolution of the zero distance FS, or Big Bang, metric of $\Gr(\BC^{n+1})$ toward other available positive definite metrics.
The unwrapping of the very tight folds of the Big Bang metric corresponds to the liberation of an ever greater number of degrees of freedom, the expected increase in entropy with time.
We can also conceive that such a dynamical evolution could lead ultimately to a rejamming, a ``mini-Big Crunch,'' which is the end result of gravitational collapse.
This last event proceeds as a kind of time-reversed counterpart of our resolution of the Big Bang low entropy problem.
The degrees of freedom holographically encoded in the black hole horizon are thus enfolded in the very structure of the complex non-linear Grassmannian.

The properties of $\Gr(\BC^{n+1})$ suggest that it has the potential to reveal the nature of quantum gravitational systems whose descriptions have long remained elusive.
The connection between this space and freezing by heating systems tells us that our Universe evolved away from an initial jammed state.
This allowed galaxies to form and life to evolve, so that we could begin rejamming the Universe all over again, one car at a time.

\vskip.5cm
\paragraph{Acknowledgments:}
CHT thanks Cornelia Vizman for first alerting him to the key paper by Michor and Mumford on vanishing geodesic distance and Mathieu Molitor for many mathematical clarifications on non-linear Grassmannians.
VJ expresses gratitude to the LPTHE, Jussieu for generous hospitality.
DM \& MK are supported in part by the U.S.\ Department of Energy under contract DE-FG05-92ER40677.

{\small

\end{document}